\documentclass[aip,rsi,reprint]{revtex4-1}

\usepackage{graphicx}
\usepackage{dcolumn}
\usepackage{bm}
\usepackage{url}
\usepackage{color}

\usepackage{amssymb} 

\begin{document}



\title{Magnetic cooling for microkelvin nanoelectronics on a cryofree platform} 



\author{M. Palma}
\email[]{These authors contributed equally to this work.}
\affiliation{Department of Physics, University of Basel, Klingelbergstrasse 82, CH-4056 Basel, Switzerland}

\author{D. Maradan}
\email[]{These authors contributed equally to this work.}
\affiliation{Department of Physics, University of Basel, Klingelbergstrasse 82, CH-4056 Basel, Switzerland}
\affiliation{Physikalisch-Technische Bundesanstalt~(PTB), Bundesallee 100, 38116 Braunschweig, Germany}

\author{L. Casparis}
\affiliation{Department of Physics, University of Basel, Klingelbergstrasse 82, CH-4056 Basel, Switzerland}
\affiliation{Center for Quantum Devices, Niels Bohr Institute, University of Copenhagen, 2100 Copenhagen, Denmark}

\author{T.-M. Liu}
\affiliation{Department of Physics, University of Basel, Klingelbergstrasse 82, CH-4056 Basel, Switzerland}
\affiliation{Department of Applied Physics, National Pingtung University, Pingtung City, Taiwan}

\author{F. Froning}
\affiliation{Department of Physics, University of Basel, Klingelbergstrasse 82, CH-4056 Basel, Switzerland}

\author{D. M. Zumb\"{u}hl}
\email[]{dominik.zumbuhl@unibas.ch}
\affiliation{Department of Physics, University of Basel, Klingelbergstrasse 82, CH-4056 Basel, Switzerland}

\date{\today}

\begin{abstract}
We present a parallel network of 16~demagnetization refrigerators mounted on a cryofree dilution refrigerator aimed to cool nanoelectronic devices to sub-millikelvin temperatures. To measure the refrigerator temperature, the thermal motion of electrons in a Ag~wire -- thermalized by a spot-weld to one of the Cu nuclear refrigerators -- is inductively picked-up by a superconducting gradiometer and amplified by a SQUID mounted at 4\,K. The noise thermometer as well as other thermometers are used to characterize the performance of the system, finding magnetic field independent heat-leaks of a few nW/mol, cold times of several days below 1\,mK, and a lowest temperature of 150$\,\mu$K of one of the nuclear stages in a final field of 80\,mT, close to the intrinsic SQUID noise of about 100$\mathrm{\,\mu K}$. A simple thermal model of the system capturing the nuclear refrigerator, heat leaks, as well as thermal and Korringa links describes the main features very well, including rather high refrigerator efficiencies typically above 80\,$\%$.
\end{abstract}


\maketitle 

\section{Introduction}
As thermal excitations represent an ubiquitous energy scale in solid state systems, advancing to lower temperatures might open up the way to the discovery of new physical phenomena such as fragile fractional quantum Hall states~\cite{pan_1999} and electron-mediated nuclear phase transitions, both in 2D and 1D systems~\cite{simon_2007,simon_2008,scheller_2014}. To investigate such phenomena, one needs to access lower temperatures beyond what a dilution refrigerator could achieve. Adiabatic Nuclear Demagnetization~(AND)~\cite{pickett,clark_2010} is a well very well established technique with the potential to open the door to the $\mu$K-regime for nanoelectronics. In many laboratories, the sample is only weakly coupled to the coldest spot of the refrigerator, resulting in sample temperatures significantly higher than the base temperature of the refrigerator. In order to efficiently couple sample and refrigerator, a parallel network of Nuclear Refrigerators~(NRs) was proposed~\cite{ clark_2010,Anna_2015}, where every lead is well thermalized through the mixing chamber~(MC) and has its own NR. Our approach relies on the Wiedemann-Franz cooling of the conduction electrons~\cite{casparis_2012}, which is the main cooling mechanism in the mK-regime and below.
\par The implementation of a parallel network of NRs on a cryogen-free system is very challenging due to the increased vibration level compared to a wet system. However, cryogen-free platforms will become more important for low temperature experiments, because they offer ample experimental space and operation without liquid helium, thus reducing costs and dependence on helium infrastructures. In addition, particularly referring to AND, cryogen-free systems are suitable for longer precooling and extended hold time compared to the traditional wet system, due to liquid He transfers increasing the temperature of the entire system. First operative AND systems on cryogen-free platforms have been implemented using both PrNi$_5$ and Cu as nuclear refrigerant~\cite{batey_2013,casey_2014,todoshchenko_2014}. In contrast to the single nuclear stage experiment, the parallel network of NRs amends itself for nanoelectronics providing direct cooling of the electrons in each of the wires connected to the sample.
\par In this article, we present a successful implementation of a parallel network of Cu NRs on a cryogen-free platform demonstrating cooling with high efficiency close to ideal adiabatic behavior down to 150\,$\mu$K. The temperature is measured using an inductive Johnson noise thermometer~\cite{engert_2007,engert_2012,beyer_2007}, which operates over a broad range of temperatures from 4\,K down to 150\,$\mu$K. The noise thermometer is an ideal choice for low temperature applications, because self-heating is reduced due to the inductive read-out and the thermometer has the potential to reach the low $\mu$K-regime. We measure field independent heat leaks of less than 2\,nW/mol for magnetic fields below 1T, allowing the NRs to stay below 1\,mK for roughly 50 hours at 80 mT. We model the AND process and obtain a dynamic heat leak independent of the magnetic field ramp rate. Thus, it is possible to increase the efficiency of the AND process by reducing the duration of the $B$-field ramping.

\section[Nuclear Refrigerator Network on a Cryofree Platform]{Nuclear Refrigerator Network on a Cryogen-Free Platform}\subsectionmark{Nuclear Refrigerator Network}
Recently, AND experiments have been successfully implemented on a cryogen-free platform, using PrNi$_5$ and Cu as nuclear refrigerant~\cite{batey_2013,casey_2014,todoshchenko_2014}. With PrNi$_5$, reaching ultra-low temperatures is restricted to the rather high nuclear ordering temperature~($T\,\sim$\,0.5\,mK)~\cite{pobell,Mueller_1980}.
In contrast, Cu can be demagnetized down to the low $\mu$K regime and it is very easy to work with and to machine, particularly compared to PrNi$_5$. The high electrical conductivity of Cu makes it susceptible to eddy current heating, which arises from both ramping of the magnetic field and vibrations in a non-homogenous $B$-field. The pulse tube~(PT) cold head is a powerful source of both cooling and vibrations, making the implementation of AND an exacting task. Adiabatic nuclear demagnetization experiments are very susceptible to heat leaks, increasing temperature and accelerating the warm up of the NRs, thus reducing the hold time. The concept of a parallel network of 16 Cu NRs presented here overcomes these challenges and leads to a straightforward integration of the AND technique into transport measurement setups.

\begin{figure}[t]
	\centering
	\includegraphics[width=8.7cm]{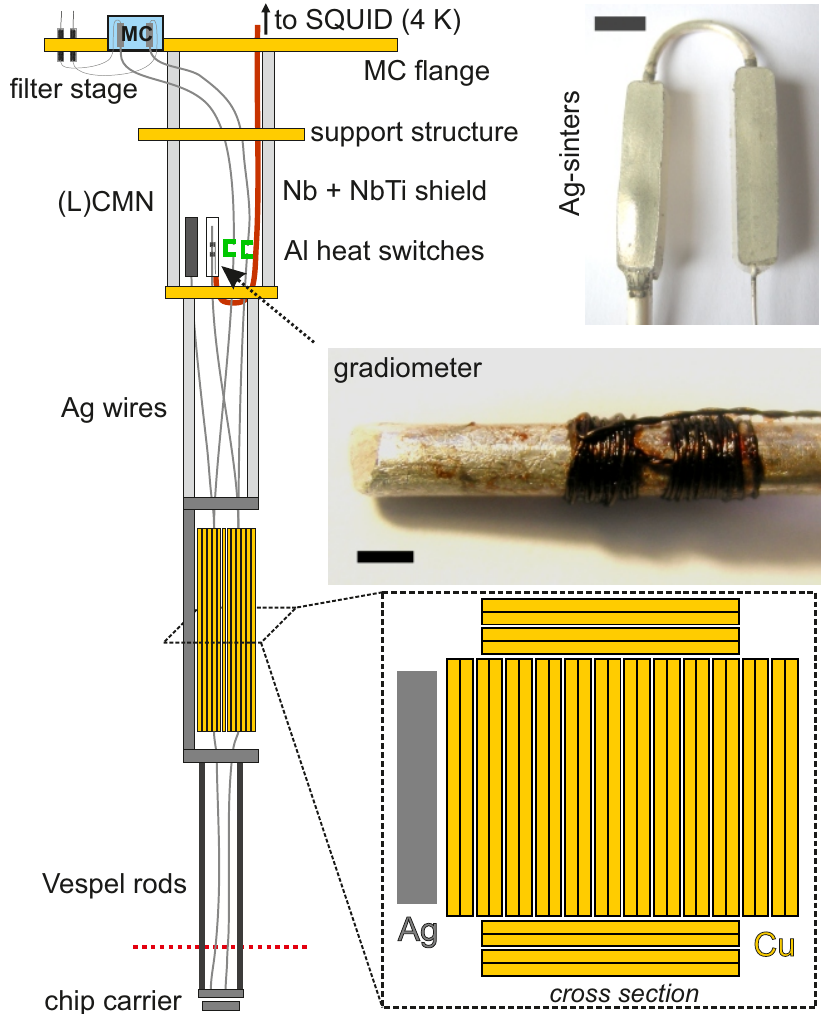}
	\caption{Schematic of the nuclear demagnetization stage. The measurement leads are thermalized with Ag powder sinters~(\emph{top right picture}, scale bar:~5\,mm) in the mixing chamber~(MC, blue) and pass through C-shaped Al heat switches~(green) to the Cu plates. The gradiometer of the noise thermometer as well as the~(L)CMN thermometers are positioned in a region of cancelled magnetic field between the MC and the NR stage. The gradiometer is double-shielded by a Nb tube and a outer NbTi tube~(red). \emph{Middle right inset:}~photograph of the gradiometer pick-up coil made from insulated Nb wire with $100\,\mu$m diameter. The 2x\,20 turns are wound non-inductively  on a high-purity silver wire which is spot-welded to a NR. Scale bar:~2\,mm. \emph{Lower inset:}~schematic cross section through the network of 16~parallel NRs.}
	\label{fig:BFscheme}
\end{figure}

Figure~\ref{fig:BFscheme} shows a schematic of the nuclear stage. Starting from the top, the measurement leads are filtered by lossy thermocoax~\cite{zorin_1995} from room temperature~(RT) to the MC flange of the dilution refrigerator. Additional filtering is achieved by home-built Ag-epoxy filters~\cite{schellerfilter_2014} and double-stage RC filters bolted to the MC flange. Next, each of the 16~leads is thermalized inside the Cu MC using Ag powder sinters, as shown in the top right inset of Fig.~\ref{fig:BFscheme}, which are electrically insulated from the MC~(ground) and each other. To allow the passage of the leads through the MC we designed super-fluid leak-tight feedthroughs on the bottom of the MC. The leads exit the MC as annealed Ag-wires, then pass through the Al heat switches with fused joints~\cite{lawson_1982} and finally they are spot-welded to Cu plates. At the bottom of each NR, another annealed Ag wire continues to the chip carrier, providing a platform for nanostructured samples on an easily exchangeable chip carrier, see Fig.~\ref{fig:BFscheme}. Therefore, each lead provides a thermally highly conductive path from the sample to the NR, electrically insulated from all other wires and ground. The chip socket below the red dashed line in Fig.~\ref{fig:BFscheme} was not mounted during the measurements in the main text, but can easily be added without significant influence on refrigerator performace. For additional details about the measurement setup see the supplementary materials~\cite{supplementary}.
\par Magnetic fields up to 9\,T can be separately and independently applied to the AND stage and the sample. The C-shaped Al pieces are used to implement the concept of heat switches allowing to choose between excellent or very poor thermal conductivity, while always keeping the sample electrically connected. In the superconducting state Al is a thermal insulator while in its normal state, when the superconductivity is broken by a small magnetic field~($\geq$10\,mT), it is an excellent thermal conductor.  All the thermometers used in the experiment are susceptible to magnetic fields; therefore they are positioned together with the Al heat switches in a region of canceled magnetic field between the MC and the NRs and are double shielded by Nb and NbTi tubes. The three thermometers in use are a Cerium Magnesium Nitrate~(CMN) thermometer, a Lanthanum diluted CMN~(LCMN) thermometer and the Johnson noise thermometer. Each thermometer is connected to its own NR through a massive Ag wire of 25 cm length.

\par Although there are no mechanically moving parts in state-of-the-art pulse tubes, vibrations caused by high-pressure gas oscillations and the compressor package are transduced to the refrigerator. Despite significant progress in recent years, cryogen-free systems tend to have drastically increased vibration levels compared to standard systems~(i.e.\ dewars with cryo liquids). To account for these challenges special care was taken on damping all connections to the fridge and decoupling the PT cold head~\cite{bluefors} from the rest of the system. The presented setup was improved from a previous wet system~\cite{clark_2010,Anna_2015} to particularly meet the demands of a cryogen-free system~\cite{bluefors}. We introduced a rigid support structure and an adapted geometry of the NRs. Compared to the wet system version~\cite{casparis_2012}, we decreased the cross section relevant for eddy current heating and simultaneously doubled the amount of Cu per plate. Further, the surface area of the Ag-sinters was tripled to now 9\,$\rm{m}^2$ per lead and the diameter of the Ag wires is doubled, since these thermal resistances have been identified as a bottle neck during precooling~\cite{clark_2010}.

\begin{figure}
	\centering
	\includegraphics[width=8.7cm]{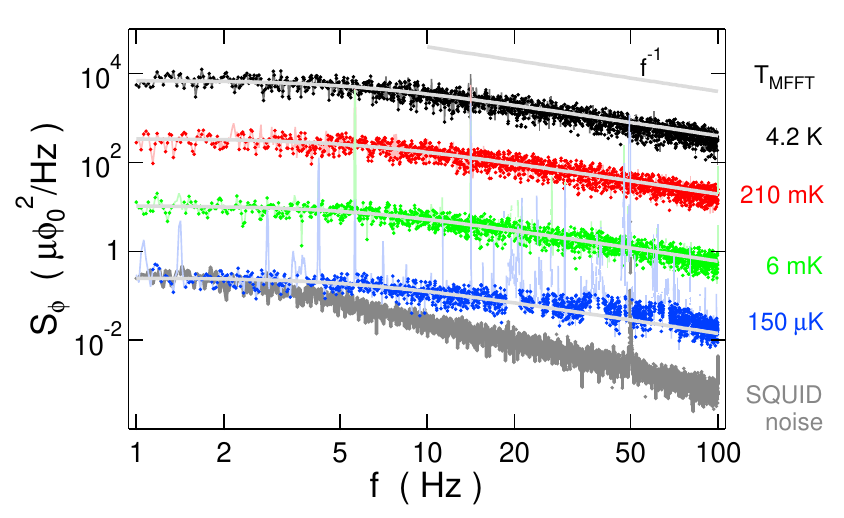}
	\caption{ Power spectral density $S_{\phi}(0,T_{\rm{MFFT}})$ of the magnetic flux noise, in units of the flux quantum $\phi_{0}$, at various NR temperatures. The light gray solid curves are fits using Eq.~(\ref{eq:sphifit}) which are converted to $T_{\rm{MFFT}}$ as described in the text using the reference spectrum at $T_{\rm{ref}} = 4.2$\,K. The noise peaks become more visible at lower temperatures where the thermal noise becomes smaller. The SQUID noise shown here in grey is from a similar SQUID with the inputs shorted, not from the SQUID used to measure the MFFT.}
	\label{fig:spectra}
\end{figure}

\section{Noise Thermometry}
Measuring temperatures in the $\mu$K regime is a challenging task, since many thermometers suffer from self-heating and are very  susceptible to heat leaks, often leading to a saturation of the thermometer. Here we use a specific type of noise thermometer, the magnetic field fluctuation thermometer~(MFFT), based on an inductive read-out~\cite{rothfuss_2013,engert_2007,beyer_2007,engert_2012} using a SQUID amplifier, which is designed to reduce internal and external heat leaks. In fact, we reduce the external heat leaks onto the thermometer by mechanical and electrical decoupling of shield and thermal noise source, see supplementary material. In addition, thanks to the inductive read-out, the thermal noise is detected without any bias applied, thus reducing self-heating of the thermometer. Our improvements allow us to measure temperatures down to 150 $\mu$K, whereas the lowest measurable temperature is roughly 100 $\mu$K given by the SQUID noise floor. Figure~\ref{fig:spectra} shows the spectra for various temperatures. Note that for the lowest temperature~(blue) the spectrum is just above the SQUID noise spectrum~(grey).

The temperature of electrons in a metal is related to their thermal~(Brownian) motion, which generates current noise given by the Johnson-Nyquist formula~\cite{Johnson,Nyquist}:

\begin{equation}
\label{eq:Johnson-Nyquist}
S_{I}=\frac{4k_{B}T}{R} \,.
\end{equation}

Here, $S_{I}$ is the power spectral density of the current  noise, $k_{B}$ is the Boltzmann constant, $T$ is the temperature of the electrons in the conductor and $R$ is the resistance of the metal. The read-out of the thermal noise is done by a gradiometer, consisting of two counter-wound superconducting pick-up coils~(detector) wrapped tightly around a Ag wire~(thermal noise source) of radius $r$. The working principle is the following: thermal currents are transformed, by self-inductance, into magnetic flux fluctuations, detected by the pick-up coil. In the low frequency range, the power spectral density of the magnetic flux noise~\cite{rothfuss_2013,varpula_1984} can  be written as $S_{\phi}(0,T_{\rm{MFFT}})=4k_{B}T_{\rm{MFFT}}\sigma G\mu_{0}^2r^{3}/2\pi$, where $\mu_{0}$ is the vacuum permeability, $T_{\rm{MFFT}}$ is the electronic temperature measured with the  MFFT, $G$ is a geometric factor~\cite{rothfuss_2013} and $\sigma$ is the electrical conductivity at low temperatures, which is assumed temperature independent in the mK range. The electrical conductivity is defined as $\sigma=\sigma_{\rm{RT}}*RRR$, with RRR being the residual resistivity ratio and $\sigma_{\rm{RT}}$ the room temperature conductivity. Note that $S_{\phi}(0,T_{\rm{MFFT}})$ depends linearly on temperature $T_{\rm{MFFT}}$, since $\sigma$ is constant at low frequency where the Skin effect is negligible. Figure~\ref{fig:spectra} shows that all the spectra have a low frequency plateau.

The Skin effect forces high frequency current fluctuating to the metal surface. As a consequence the conductivity of the Ag wire becomes frequency dependent, resulting in a low-pass like shape of $S_{\phi}(f,T_{\rm{MFFT}})$.  Such a frequency dependence is described by the following equation:

\begin{equation}
\label{eq:sphifit}
S_{\phi}(f,T_{\rm{MFFT}})=\frac{ S_{\phi}(0,T_{\rm{MFFT}})}{\sqrt{ 1+\left(\frac{f}{f_c}\right)^2 }} \,,
\end{equation}
 where the cut-off frequency is given by $f_c = 4.5/(\pi\mu_0\sigma r^2)$~\cite{rothfuss_2013}. As expected in Fig.~\ref{fig:spectra} the amplitudes of the spectra  decrease as $1/f$ at high frequencies.

In order to measure\footnote{The DC-SQUID is operated in a flux-locked loop mode with XXF-1 electronics~\cite{magnicon}, including a second order Bessel-type low-pass filter~($f_{\rm{3dB}} = 10$\,kHz). After a room temperature voltage preamplifier with another low-pass filter~($f_{\rm{3dB}} = 1$\,kHz), the signal is acquired with a digital-to-analog converter.} the power spectral density, we need to acquire 10 real-time noise traces with 50\,s duration each, which are averaged after Fourier transformation.  Noise peaks appear mostly at frequencies corresponding to higher harmonics of the rotation frequency of the PT motor~(1.4\,Hz), see light blue spectrum in Fig.~\ref{fig:spectra}. The peaks become more evident at low temperatures due to the lower thermal noise background. To eliminate these peaks we fit every spectrum first with a polynomial of 10\textsuperscript{th} order and eliminate any data that exceeds the polynomial by more than an empirically determined threshold factor.

The MFFT is used as a secondary thermometer calibrated against the MC thermometer at 4.2\,K. From the fit of Eq.\,(2) to the reference spectrum at 4.2\,K, we extract a value of $f_{c}\sim5$\,Hz independent of temperature and corresponding  to a RRR of about 2000, which matches typical RRRs determined in independent transport measurements. Thus, we fix $f_{c}=5$\,Hz and extract $T_{\rm{MFFT}}$ as the only fit parameter from the fit to Eq.~(\ref{eq:sphifit}).

To achieve optimal performance of the MFFT, we addressed and solved the following technical issues. First, to avoid effects from homogeneous magnetic fields, the pick-up coil is non-inductively wound around the Ag wire. Second, the gradiometer shown in Fig.~\ref{fig:BFscheme} is connected by a long section of twisted Nb wires to the SQUID, which is anchored to the quasi-4\,K-flange of the refrigerator. The twisted Nb wires are double-shielded with a Nb and NbTi tube, both thermalized at the mixing chamber plate, cold plate and still plate. Finally, we mounted the SQUID at 4\,K to avoid low-frequency excess flux noise~\cite{wellstood_1987}, which can arise at sub-K temperatures.


\section{Nuclear Refrigerator Performance}
The nuclear refrigerator technique is based on a single shot cycle consisting of the following steps: magnetization in an initial field $B_{\rm{i}}$=9\,T, precooling  down to $T_{\rm{i}}$=10\,mK~(three days), and demagnetization down to a final field $B_{\rm{f}}$. Finally, low temperatures can be explored over a period of time while the system continuously warms up, due to a small parasitic heat leak~($\dot{Q}$) absorbed by the NRs. During the AND process the nuclear temperature of the Cu plate is lowered from the initial temperature  $T_{i}$ down to the final temperature $T_{f}$. During magnetization and precooling, the Al heat switches are set to conduct heat excellently~(normal state) to cool the NRs via the MC. While demagnetizing and warming up, the heat switches are superconducting to prevent heat flowing from the MC into the NRs. In this refrigeration technique, the nuclear spin degree of freedom has by far the largest heat capacity, absorbing the heat leaks coming into the NRs. This can lead to non-equilibrium configurations where other degrees of freedom~(e.g. electrons, phonons) can be at different temperatures than the nuclear spins, due to the finite thermal conductivity between them.

We need to characterize the heat leak of the system, which then defines the efficiency $\xi=(T_{\rm{i}}/T_{\rm{f}})/(B_{\rm{i}}/B_{\rm{f}})$  of the AND process.  An efficiency of 100\,\% indicates a fully adiabatic and reversible process while $\xi$ less than 100\,\% signifies the presence of heat leaks, which spoil the adiabaticity of the AND process. One distinguishes two types of heat leaks: a static heat leak~($\dot{Q}$) appearing already at fixed magnetic field and attributed mainly to heat release, radiation and vibrations. Beyond that, an additional dynamic heat leak~($\dot{Q}_{\rm{dyn}}$) appears when sweeping the magnetic field.

\begin{figure}
	\centering
	\includegraphics[width=8.7cm]{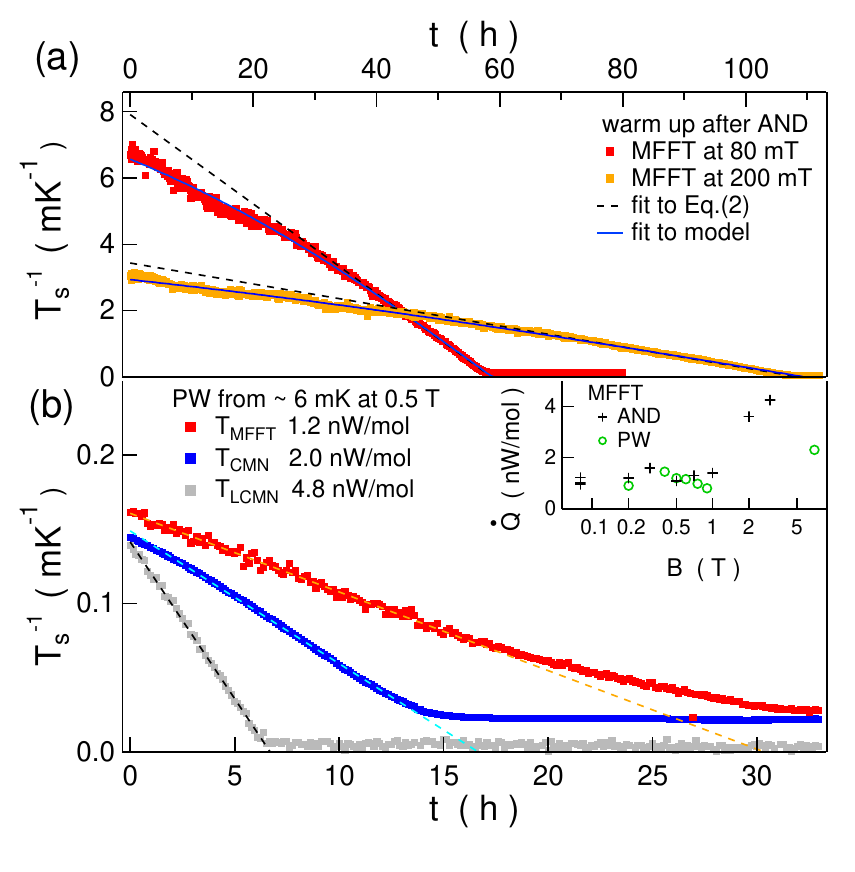}
	\caption{Warm-up curves: (a) Inverse of  $T_{\rm{s}}$ measured with the MFFT versus time during the warm up at 80\,mT~(red squares) and 200\,mT~(orange squares) after AND. The linear fits~(black dashed lines) reveal extrapolated electron temperatures $T_{\rm{ex}}$\,=\,126/280~$\mu$K at the beginning of the warm up and heat leaks of $\dot{Q}$~=~0.9/1.16~nW/mol for  80\,mT  and 200\,mT, respectively. The solid blue curves are the fits to the heat flow model~(see main text) with $\dot{Q}_{\rm{s}}$\,=\,6\,\% at 80\,mT and 18\,\% at 200\,mT of the total static heat leak.  (b) Precool and warm up (PW) measurements: $T_{\rm{s}}^{\rm{-1}}$ during warm up, from MFFT, CMN and LCMN thermometers versus the time after opening the heat switches at B\,=\,0.5\,T, resulting in $\dot{Q}$\,=\,1.2/2.0/4.8~nW/mol, respectively. Inset: static heat leak $\dot{Q}$ to the nuclear stage per mol of Cu, measured with the MFFT at various $B$-fields, extracted using Eq.~(\ref{eq:warmupfit}) after AND (black crosses) and after PW (green circles).
	}
	\label{fig:heatleaks}
\end{figure}

To determine the static heat leak onto a NR, we read a sensor temperature $T_{\rm{s}}$ as a function of time during the warm up, see Fig.~\ref{fig:heatleaks}, displaying $T^{-1}_{\rm{s}}$. Our sensor cannot operate directly on the NR due to the magnetic fields present, and thus is placed at some distance and is thermally well connected to the NR through a high-conductivity Ag wire. Over time, the temperature is continuously increasing until it saturates at rather high temperature $\sim50$\,mK, far exceeding the MC temperature $T_{\rm{MC}}\,\sim$\,7\,mK. At this point, the heat leaking from the NR through the Al heat switches into the MC balances the static heat leak, keeping the Cu stage at a constant temperature. One can model the warm up behavior of the NRs\cite{pobell,pickett} by assuming a constant static heat leak $\dot{Q}$ flowing entirely into the Cu nuclear spins:

\begin{equation}
\label{eq:warmupfit}
T_{\rm{e,Cu}}^{-1}(t) = T_{\rm{ex}}^{-1}-t\left(\frac{\lambda_n B_f^2}{\mu_0\dot{Q}} + \kappa \right)^{-1},
\end{equation}

\noindent where $T_{\rm{e,Cu}}$ is the electronic temperature of the Cu plate, $T_{\rm{ex}}$ is the extrapolated electronic temperature of the Cu plate at the beginning of the warm up, $\mu_{0}$ is the vacuum permeability, $\lambda_n$ is the molar nuclear Curie constant of Cu and $\kappa$ is the Korringa constant~\cite{pickett} for Cu. The Korringa constant quantifies thermal coupling and thus the temperature gradient between the electrons and the nuclei. As Eq.~(\ref{eq:warmupfit}) shows, $T_{e,Cu}^{-1}$ is an affine function of time. In the intermediate temperature regime, but away from saturation, we fit Eq.~(\ref{eq:warmupfit}) to the data, shown as dashed lines in Fig.~\ref{fig:heatleaks}(a). The fits are in very good agreement with the data for intermediate temperatures, which indicates that the heat leak is constant over a long period of time.
\par From the fit we extract $\dot{Q}$ and $T_{\rm{ex}}$. The black crosses in the inset of Fig.~\ref{fig:heatleaks}(b) show the measured $\dot{Q}$ for ANDs at various final fields. As seen, $\dot{Q}$ is roughly 1\,nW/mol and independent of $B_{f}$ below 1\,T. This is striking since it indicates negligible eddy current heating. During a warm up, the magnetic field is held constant but eddy current heating could still arise due to vibrations in an inhomogeneous magnetic field: $\dot{Q}\propto (dB/dt)^2 = [(dB/d\mathbf{r})(d\mathbf{r}/dt)]^2$.
\par As shown in Fig.~\ref{fig:heatleaks}(a), the temperature sensor shows a saturation in the low temperature regime and lies below the theory curve of inverse temperature. Such an elevated sensor temperature $T_{\rm{s}}$ can be caused by heat release, e.g. at the thermometer itself. The temperature gradient between the sensor and the NR can be taken into account using a heat flow equation, that fits the sensor temperature of the MFFT in the whole dynamic range. The total static heat leak can be decomposed into a sensor heat leak $\dot{Q}_{\rm{s}}$ and a remaining heat leak directly acting onto the NRs. The temperature gradient due to the heat leak $\dot{Q}_{\rm{s}}$ can be written as :

\begin{equation}
\label{eq:heatflow}
T^{2}_{\rm{s}}(t)-T_{\rm{e,Cu}}^{2}(t) = \frac{2}{\kappa_{0}}\dot{Q}_{\rm{s}} .
\end{equation}

The difference between the square of the two temperatures comes from the integration of the thermal conductivity of the metallic link between the sensor and the NR, which is linear in T. The coefficient $\kappa_{0} =\pi^{2}k_B^{2}/3e^{2}R_{\rm{tot}}$, where $e$ is the electron charge, depends on the total resistance  $R_{\rm{tot}}$\,$\sim$\,1\,$\mu\Omega$ comprised in similar parts from the spot welded junction between the Ag wire and the Cu plate and the resistance of the Ag wire. Note that the low temperature resistivity is reduced by a RRR\,$\sim$\,2000, achieved by annealing the high purity Ag wire. By plugging Eq.~(\ref{eq:warmupfit}) into Eq.~(\ref{eq:heatflow}), we obtain $T_{\rm{s}}^{\rm{-1}}$ as function of the time with $\dot{Q}_{\rm{s}}$ as an additional fit parameter. The solid blue curves in Fig.~\ref{fig:heatleaks}(a) show the best fit, exhibiting excellent agreement down to the lowest temperatures. The sensor heat leak $\dot{Q}_{\rm{s}}$ is between 5 and 20\,\% of $\dot{Q}$, indicating a rather small heat leak emanating from the MFFT.

Performing a complete AND experiment to extract $\dot{Q}$ for different $B$-fields is very time-consuming. In order to procure $\dot{Q}$ faster, we introduce an abbreviated precool and warm up~(PW) cycle: The Cu stage is precooled at fixed magnetic field and subsequently warms up due to the heat leak $\dot{Q}$, after being thermally isolated from the MC with the heat switch. Figure~\ref{fig:heatleaks}(b) shows the warm-up of a PW cycle at a $B$-field of 0.5 T for all the thermometers in use. For all three sensors, the inverse of the temperature decreases linearly in time and eventually saturates at high temperature. Note that each of the three thermometers have their own saturation temperature~(high T) and warm up time, indicating different heat leaks. By using Eq.~(\ref{eq:warmupfit})~(dashed line in Fig.~\ref{fig:heatleaks}(b)) we extract a $\dot{Q}$ for the MFFT  of around 1 nW/mol and find higher values of 2 nW/mol for the CMN and 4.8 nW/mol for the LCMN. The heat leaks extracted with PWs for different $B$-fields are consistent with the ones from warm ups after AND, see inset Fig.~\ref{fig:heatleaks}(b). 
Note that  for the MFFT a minute amount of GE Varnish is used to fix the superconducting pick-up coil to the silver wire while for the packaging of the (L)CMN, a  considerable amount of epoxy is used, which is a well-known source of heat release.


\par
\begin{figure}
	\centering
	\includegraphics[width=8.7cm]{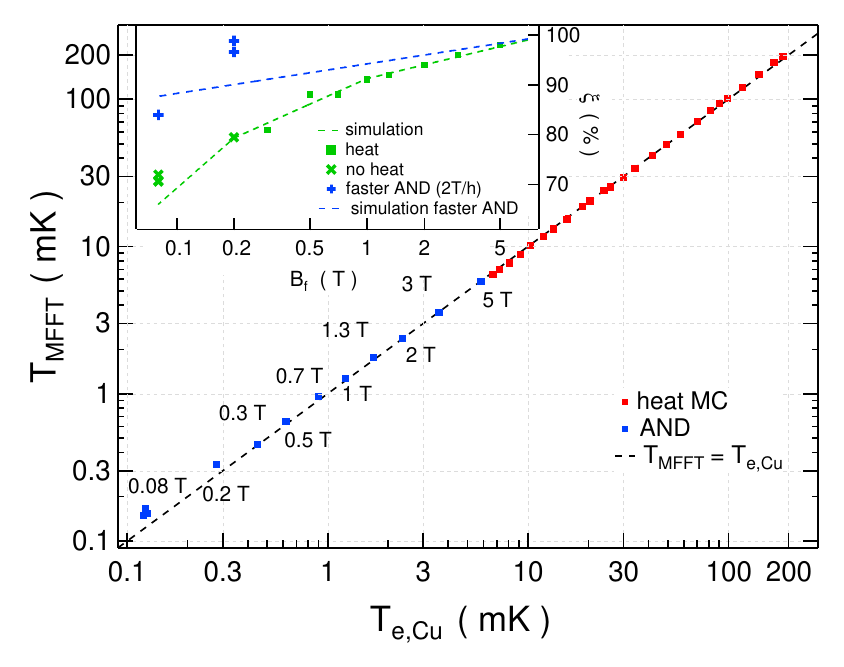}
	\caption{Temperature measured by the noise thermometer $T_{\rm{MFFT}}$ versus temperature of the nuclear stage $T_{\rm{e,Cu}}$. 
	Above 7 mK, $T_{\rm{e,Cu}}$ is measured with a calibrated RuO$_2$ thermometer sitting on the MC~(red squares). In this temperature range the MC and the Cu stage are well thermally coupled. For temperatures below 7\,mK, $T_{\rm{e,Cu}}$ is extracted from warm ups after AND~(blue squares)~(see text and Fig.~\ref{fig:heatleaks}(a)). The black dotted line represents $T_{\rm{MFFT}} = T_{\rm{e,Cu}}$. \emph{Inset:}~the efficiency $\xi$ as function of the final magnetic fields $B_{\rm{f}}$. The dashed curves show simulations of the AND process, carried out at two different ramp speeds. The markers show the efficiency extracted from different AND runs as labeled. }
	\label{fig:tcalib}
\end{figure}

 Next, we compare the electronic temperature of the Cu plates $T_{\rm{e,Cu}}$ as extrapolated from the warm up curves after AND~($T_{\rm{ex}}$) with the measured electronic temperature $T_{\rm{MFFT}}$, finding excellent agreement, as seen in Fig.~\ref{fig:tcalib}, blue squares. Even though the thermometers used for the extrapolation~(CMN and LCMN) become fully saturated at rather high temperatures, here around 2.5\,mK, the extrapolation method -- as also relied on in our previous works \cite{casparis_2012, clark_2010,Anna_2015} -- is seen here to work rather well down to the lowest temperatures measured. As shown in Fig.~\ref{fig:tcalib} below 400\,$\mu$K, $T_{\rm{MFFT}}$ starts to be slightly higher than $T_{\rm{e,Cu}}$, reaching a maximum deviation of 20\,\% for the lowest temperature. At 150 $\mu$K the MFFT is mainly limited by the SQUID noise level and hence slightly higher than the lowest extracted temperature $T_{\rm{ex}}$=120\,$\mu$K. In the high temperature regime, the MFFT is tested against a calibrated $\rm{RuO_{2}}$ thermometer sitting on the MC flange, showing excellent agreement of the temperature reading of the two thermometers, see red squares in Fig.~\ref{fig:tcalib}.

\par To complete the characterization of the AND system, we now turn to the efficiency of the process. As shown in the inset of Fig.~\ref{fig:tcalib} the efficiency decreases monotonically from almost 100\,\% at high final magnetic field down to 70\,\% for the lowest final field. The reduction of the efficiency for lower magnetic fields is a result of the smaller heat capacity of the Cu nuclei, which is proportional to $B_{\rm{f}}^2$. 
We simulated the efficiency of the AND process assuming $\dot{Q}_{\rm{dyn}}$ depending linearly or quadratic on $B$ or $\dot{B}$ as one would expect for $\dot{Q}_{\rm{dyn}}$ arising from vibration or eddy current heating, but in these cases the simulations completely missed the experimental points. In contrast, assuming a fixed $\dot{Q}_{\rm{dyn}}$ of 29\,nW/mol, independent $B$ or $\dot{B}$, reproduces the data well~(green dashed curve in inset of Fig.~\ref{fig:tcalib}). Thus, the simulation suggests that $\dot{Q}_{\rm{dyn}}$ is constant in time and independent of the ramp-rate of the $B$-field, which gives the opportunity to increase the efficiency by reducing the duration of the demagnetization process. This hypothesis was successfully tested in the experiments by doubling the ramp speed of the AND, as shown by the blue crosses in the inset, where $\xi$ increases significantly for the faster rate -- albeit the simulation predicts slightly different efficiencies than those measured.

To estimate the dynamic heat leak, we first open the switches, then we ramp the magnetic field from zero to a finite $B$-field and back to avoid any nuclear contribution to the heat capacity. We measured the temperature of the Cu plates and by integrating the electron heat capacity we obtain the energy stored in the system. In addition, we estimate the heat leaking through the superconducting Al heat switches due to phonon-dislocation scattering processes~\cite{pobell,gloos_1990}. This estimation yields $\dot{Q}_{\rm{dyn}}$ of 22 nW/mol for a ramp-rate of 1\,T/h, which is comparable to the value used in the simulation. However, the $\dot{Q}_{\rm{dyn}}$ estimated with this protocol is dependent on the ramp-rate, which is in disagreement with the simulation. Our simple model needs further work to fully understand the origin of $\dot{Q}_{\rm{dyn}}$ and its dependence on the sweep rate.

\section{Conclusions}

In summary, we have implemented a parallel network of 16~electrically separated NRs on a cryogen-free platform. These 16~plates are part of the measurement leads and can be straightforwardly used to cool nanostructured samples. The nuclear stage is equipped with a magnetic field fluctuation thermometer, showing excellent agreement with the NR temperature $T_{\rm{e,Cu}}$ down to 400\,$\mu$K. After AND to $B_f = 0.08$\,T, the lowest temperature reading is limited to $150\,\mu$K while the extrapolated electron temperatures is $120\,\mu$K, indicating good agreement between the model and measurements. The heat leak measured on the NRs is around 1\,nW/mol and allows the AND stage to stay below 1\,mK for roughly 50\,hours, see supplementary material~\cite{supplementary}. Higher $B_f$ allow for even longer hold times, while still supplying reasonably low temperatures. In addition, we characterized the dynamic heat leak, which appears to be constant in time and independent of the sweep rate of the magnetic field, making possible significantly increased efficiency at faster magnetic field sweep rates.

\begin{acknowledgments}
We would like to thank H.\,J. Barthelmess, R. Blaauwgeers, G. Pickett, M. Steinacher and P. Vorselman for useful input and discussions. The work shop team of S.~Martin is acknowledged for technical support. This work was supported by the Swiss NSF, NCCR QSIT, the Swiss Nanoscience Institute, the Europeam Microkelvin Platform, an ERC starting grant (DMZ), and EU-FP7 MICROKELVIN and SOLID.

\end{acknowledgments}


%

\end{document}